\documentclass[a4paper,11pt]{article}
\pdfoutput=1 

\usepackage{jheppub} 

\usepackage[T1]{fontenc} 
\usepackage{mathrsfs}
\usepackage{braket}
\usepackage{inputenc}
\usepackage{hyperref}

\title{\boldmath{\huge{S-duality and Chaos:}} }

\author[]{K. Shirish,}

\affiliation[]{Department of Physics, Visvesvaraya National Institute of Technology,\\Nagpur, 440010, India}

\emailAdd{k.shirishrao@gmail.com}

\abstract {The Renormalization group in field theories happens to resemble with dynamical systems in many ways. In this paper, we discuss the unexpected connection between chaos and duality in field theories.In a sense, that various dual field theories can emerge at the end of chaotic RG trajectories, and hence strong-weak duality in quantum ﬁeld theory is a direct result of the chaotic ﬂow of the renormalization group. This suggests that various properties of field and string theories could come into existence due to chaotic RG flow. We also conjecture the existence of dual quantum ﬁeld theories in the half strip of Riemann-Zeta function.}

\begin{document} 
\maketitle
\flushbottom

\section{\large{Introduction}}

Dualities between field theories, are remarkable because it relates a strongly coupled theory to a weakly coupled one, and hence is useful for evaluating a theory at strong coupling, where perturbation theory breaks down by translating it into it's dual description with a weak coupling constant therefore duality implies that a single quantum system has two classical limits. In this paper, we describe strong-weak duality in QFT as a chaotic phenomenon.\ Chaos in classical physics is a system having sensitive dependence on its initial conditions: a region of phase space experiencing deviations along unstable directions. An initially localized system of fixed volume evolves into a highly complex structure, while still maintaining the same volume as the initial patch, as required by Liouville's theorem.  \par{The quantum analog of classical chaos has been studied largely in recent times. Quantum systems that are classically chaotic are known to exhibit universal features such as Wigner-Dyson statistics for energy eigenvalues of Hamiltonian \cite{ch}, however, chaos in field theories is much harder to study. Recent insight has been to extend the application of  the out-of-time-order correlator \cite{2,3}- which gives a quantum mechanical analog of a Lyapunov exponent.} It has been proposed, (see \cite{4}), that  S-matrix $\langle{p'_{1},...,p'_{m}}|{p_{1},...,p_{n}\rangle} $ in quantum field theory,  where the $in$ state in the asymptotic past consists of $n$ particles with $d$-dimensional momenta $p_{i}$, and the $out$ state in the asymptotic future consists of $m$ particles with momenta $p'_{i}$, in the limit $n$, $m$ $\gg$ $1$, may exhibit chaos, in the sense any small variation of the individual momenta $p$ would cause a large change in the $out$ state. \par{ S.H. Shenker and D. Stanford in a recent paper \cite{5}, have shown, black holes to be the most chaotic objects in the universe. A tiny change in the $in$ state of the black hole's S-matrix causes an exponentially large change in the $out$ state. Consider for example adding an extra particle in the $in$ state. The effect of the additional particle will shift the black hole horizon slightly outward and increasing the Schwarzchild radius $r_{s}$. The subsequent outgoing quanta find themselves slightly closer to the horizon, which in turn increases the escape time by an amount $\delta{t} \ \sim \delta{r_{s}}\exp{\left(\Large\frac{2\pi}{\beta}\right)}$}, where $\beta$ is the inverse temperature of the blackhole state. \par{In this paper, we tend to show the relationship between S-duality and chaos. In section-2, by extending the application of the out-of-time-order correlator which gives a quantum mechanical analog of Lyapunov exponents- to quantum field theories with a large number of fields  and applying it to vertex operators, we propose the emerging of dual field theories out of chaotic Renormalization group (RG) flow. In the subsequent section, we will prove the same using the general action of quantum field theory, how S-duality emerges at the end of the chaotic flow of RG trajectories.} \subsection{\textbf{The \textit{C}-Theorem}} The Renormalization group (RG) flow in both quantum field theory and string theory has allowed us to study the parameters of the physical system at different energy scales \cite{6,7,8,9,10,11,12,13}.The concept of chaotic attractor has been discussed various times in quantum field theoratical RG equations and hence irreversibility must be the essential feature of the renormalization group flow. This irreversibilty has been studied in Zamolodchikov (weak) $c$-theorem \cite{18} which proposes the existence of a Lyapunov function $c(t)$ which is  monotonically decreasing along the RG flow toward the infrared (IR)   
\begin{equation}
\frac{dc(t)}{ds} =\beta^{i}(t)\frac{\partial c(t)}{\partial t^{i}}.
\end{equation}
where $\beta^{i}(t)$ encodes the dependence of coupling parameter $t^{i}$ on energy scale $\mu$.

{The strong c-theorem states that the beta function can be regarded as a vector field and hence  consequently we can interpret the renormalization flow geometrically as a vector field in the tangent bundle of  $M$. Hence in the case of two-dimensional field theories $\sigma(t)$ can be regarded as a morse function \cite{20}. In that case, the below equations become a gradient  flow between different field theories:}
\begin{equation}
\beta^{i}(t) = -H_{ij}(t)\frac{\partial\sigma(t)}{\partial t^{i}}
\end{equation} 
 with a symmetric metric, $ H_{ij} = H_{ji}$, and the \ $\sigma(t)$ is a Lyapunov function which satisfies the followimg relation:  
 \begin{equation}
\dot{\sigma} = \frac{d\sigma}{ds} =\sum{ -H_{ij}(t)\frac{\partial\sigma(t)}{\partial t^{i}}\frac{\partial\sigma(t)}{\partial t^{j}}}. 
\end{equation} \

  This implies that the RG flow is independent of the renormalization scheme and has some intrinsic geometric flavor to it. The fact that the RG equation can be written as a geometric statement was first noted by Lassig in \cite{21}  and later by Dolan \cite{22}. The idea there was to visualize the \ $\beta$-\ function as a vector field in the space of couplings. In this paper, instead  
   of investigating the geometric structure, we will directly work on effective action of field theories   to analyze the chaotic dynamics of RG flow.
\section{\textbf{{\large{Strong-weak duality}}}}
A new phenomenon that has risen to prominence is strong-weak duality, also known as S-duality. In some case, it is possible to spilt the hamiltonian in \cite{23} \begin{equation}
H = H_{0} + gH_{1}.\ 
\end{equation}
\begin{equation}
   = H'_{0} + g'H_{1}'.
\end{equation}
Hence  when the coupling $ g $   and $ g{'}$ have a relation as follows:
\begin{equation}
g{'} = \frac{1}{g}.
\end{equation}\
then as $g$ tends to infinte, $g'$ tends to zero, and the perturbation series in $g'$ becomes a precise illustration  of the theory just where the series in $g$ becomes ineffective. Phenomena that are difficult to compute in one description of the theory becomes simple in another and hence easy to compute. Systems in which their are multiple coulpling constant, therefore exhibit multiple dual representation.

\par{ Let's imagine we have a quantum system at $t=0$ with vertex operator \ $\alpha$ \ describing the coupling constant between interactions and allow the chaotic flow of RG  according to the equation}
 \begin{equation}
\left\langle \ \left(\Large\frac{d\alpha}{dt}\right)^{2}\right\rangle \ \sim \ \exp{(\lambda_{l}t)}.
\end{equation}

where \ $\lambda_{l}$ \  is a Lyapunov constant. 
 Let us suppose that the fields are defined over some inteval of time $t_{f}>t>t_{i}$ with initial and final boundary conditions, in this way the path integral defines a transition amplitude. However our central point in this paper is to compute the  double commutator  of the coupling constant from $t=0$ \ to \ $t=t'$  in thermal  equilibrium at inverse temperature $\beta$. \
We then have-
\begin{equation}
C(t)    =  \braket{\beta|-[\alpha_{1}(0) \ ,  \alpha(t) ]^{2}|\beta}.  
\end{equation}
 The new variable  $\alpha_{1}$  is the coupling constant evolved due to the backward chaotic RG flow. In this paper, we tend to show the relationship between  $\alpha_{1}$ \  and $\alpha$ \ as reciprocals i.e
 
\begin{equation}
\alpha_{1}  =  \frac{1}{\alpha}.
\end{equation}

\subsection{\textbf{\large{Effective field theory}}}
A simple example of duality will seem to emerge out of chaotic RG flow if we  consider a field theory with effective action
\begin{equation}
S{(\phi)} = \sum_{k}{t_k \phi^{k}}. 
\end{equation} 
which flows as
\begin{equation}
\dot{S}{(\phi)} = \chi{(\phi)} = \sum_{k}{f_k\phi^{k}}.
\end{equation}
We now define a vacuum state  $\phi_{0}$   as a minimum of $\dot{S}(\phi) = 0$.
Once we know for the RG flow of the vaccum, new situations comes into existence. For example, if we consider the potential \cite{50} :
\begin{equation}
V(\phi) = - \frac{m^{2}}{2}\phi^{2} + \frac{\lambda}{4}\phi^{4}.
\end{equation}
the quartic term in the potential is a type of self-interaction in a quantum field theory. The ground state \ $\phi_{0}$  \    changes accordingly due to renormalization of \ $m^{2}$  and \ $\lambda$  . Now  assume  that\ $\phi$ \ is further coupled to \ $\chi$ in the form 
$\chi^{4}\cos{(\omega\phi).}$ 
Then the effective coupling in the new field configuration is equal to 
\begin{equation}
g_{4}  =  \cos{(\omega\phi_{0})} = \cos{\left(\omega\sqrt{\frac{m^{2}}{\lambda}}\right).}
\end{equation}
and, for large frequency, this coupling can be oscillating along the RG flow with allowed changes for $m$ and $\lambda$.
The lowest energy minimum is a complex function of these couplings and thus the $v.e.v$ of $\phi$ would exhibit chaotic behavior . As a result of this the coupling constant would also exhibit such chaotic and irregular behavior.

Let us consider the finite temperature field theories which will exhibits the same phenomenon. By doing the wick rotation of flow time $s$, quantum field theory in $(D+1)$-dimensional spacetime is nothing but quantum statistical mechanics in $D$-dimensional space \cite{24}, therefore we could substitute the flow time $s$ by the inverse temperature
\begin{equation}  s' = is = \frac{1}{T}. 
\end{equation} 
Here $s$ represents the physical time, and $i$ is the imaginary unit. We now consider an unorthodoxical set of harmonic oscillators, with potentials. A related argument was used in \cite{25}, along with the claimed resolution in the context of chaotic RG flows:
\begin{equation}
\tilde{\alpha_{j}} +  \frac{1}{2}\omega^{2}_{j}q^{2}_{j}.
\end{equation}
with ground state energy absorbed into the intercept  $\alpha_{j}$ =$\tilde{ \alpha_{j}}$ + $\frac{1}{2}{\omega^{2}_{j}}q^{2}_{j}$. The partition function  $Z(s)$ analogous to quantum field theory under wick rotation is 
\begin{equation}
Z(s') = \exp{(S(\phi))} = \sum_{j}{\frac{\exp{(-\alpha_{j}\beta)}}{1-\exp{(-\omega_{j}\beta)}}}.  
\end{equation} \
if \  $\alpha_{j}$ $\gg$  $\omega_{j}$,  this can be approximated by  
\begin{equation}
Z(s')  =  \frac{1}{s}\sum_{\omega}{\frac{\exp{(-\alpha(\omega)\beta)}}{\omega}}.
\end{equation} 
 If we consider only two frequencies,  \    $\omega_{1}$ $\ll$  $\omega_{2}$  with   $\alpha_{1}$ $\gg$ $\alpha_{2}$, we then have

\begin{equation}
s' Z(s')  =   \frac{\exp{(-\alpha_{1}\beta)}}{\omega_{1}}  +  \frac{\exp{(-\alpha_{2}\beta)}}{\omega_{2}}.  
\end{equation} 
\begin{center}
    
for $\beta$ $\ll$ $\frac{1}{\alpha_{1}}$ \
and \ $\frac{1}{\alpha_{1}}$ $\ll$ $s$ $\ll$ $\frac{1}{\alpha_{2}}$ .
\end{center} 
Thus for many chaotically distributed coupling constant, the partition also exhibits chaotic behavior. if we consider a somewhat odd distribution of oscillators with 
$\omega_{n}  \approx  \sqrt{n} \ \  \text{and}\ \
  \alpha_{n}  =  \alpha\log{n}$. Then:

\begin{equation}
s'Z(s')  =   \zeta{\left(\frac{1}{2} + \alpha \beta\right)}  
\end{equation} 
is the Riemann-Zeta function. In imaginary time coordinate, we have
\begin{equation}
s'Z(s')  =    \zeta{\left(\frac{1}{2} +i \alpha s\right)} = 0. 
\end{equation}
Therefore from equation (2.16), we have
\begin{equation}
\alpha_{1} = - k\alpha_{2}.
\end{equation}
The existence of dualities in quantum field theories implies that the only relation these coupling constant could have is
\begin{equation}
\alpha_{1} = -k\alpha_{2}  = K \frac{1}{\alpha_{1}}.
\end{equation}
The main contribution of Jussi Lindgren and Jukka Liukkonen  was to explain the existense of imaginary unit  in quantum mechanics \cite{26}. They derived the Stueckelberg covariant wave equation from first principles via a stochastic control scheme, and from them deduced the equations of quantum mechanics. In the large N limit where QFT approaches  classical limit \cite{27}, extending the stochastic understanding of quantum mechanics to field theories we have shown, chaotic RG flow despite having sensitive dependence to initial conditions actually approaches to a dual description of the same QFT, implying strong-weak duality to be a chaotic phenomenon.

\section{\large{Conclusion and Discussions}}
We have shown that chaotic renormalization flow not only has desirable applications e.g in neural networks \cite{28} to spin glasses and especially in the cosmology of early universe as prescribed by many, and even in many field theories where various model emerges at the end of its chaotic RG trajectories, but rather at finite temperature these chaotic flow instead of ending in a big mess always result in duality. In other words, the equivalence between distinct quantum systems could be approached via chaotic RG flow. Indubitably the following argument presents an interesting connection of quantum field theories with number theory. Previous work has focused on the relationship between particular values of Riemann-Zeta function and scattering amplitudes. The concrete description comes from the work of Francis Brown, the mathematical quantities which occur as Feynmann amplitudes include values of the Riemann-Zeta function \cite{29}. However in this paper, we have added a new entry to the dictionary, assuming the trueness of the Riemann hypothesis and the chaotic RG flow, we have prescribed that all dual QFT lies in the half strip of the Riemann-Zeta function.  We suspect that we are just scratching the surface of a general connection between S-duality and chaos. ADS/CFT correspondence proposed by Juan maldacena provides a powerful framework to study strongly coupled field theories using weakly coupled gravity. As a result, it has been shown in \cite{100} that the dual of a $2D$-JT gravity in boundry theory is not a quantum system with a particular hamiltonian but rather a random ensemble of quantum systems. This seems to be quit analogous with strong-weak duality of quantum field theories in large $N$ limit in two dimensions. However the observation of the previous section in the context of ADS/CFT Correspondence is still unknown and is the need of the future research. Chaos is implicitly to play a crucial role in the description of quantum gravity, and we suspect gravity is likely to emerge out of  chaos at the end of RG trajectories. 
\acknowledgments
I am grateful to many mentors over the years, for providing many insights over the subject. I would like to thank my Professors R. S. Gedem and especially Udaysinh T. Bhosale for very useful correspondence and  their comments on a draft of this paper.
\

\bibliographystyle{unsrt}
\bibliography{biblography.bib}


\end{document}